# The Engineering of Software-Defined Quantum Key Distribution Networks

**A. Aguado[1], V. López[2], D. López[2], M. Peev[3], A. Poppe[3], A. Pastor[2], J. Folgueira[2] and V. Martín[1].**

[1]*Center for Computational Simulation, Universidad Politécnica de Madrid 28660 Madrid, Spain*

[2]*Telefonica Investigacion y Desarrollo/gCTIO, Ronda de la Comunicacion s/n 28050 Madrid. Spain*

[3]*Huawei Technologies Duesseldorf GmbH, Riesstrasse 25, 80992 Munchen. Germany*

*e-mail: a.aguadom@fi.upm.es, vicente@fi.upm.es*

***Abstract*— Quantum computers will change the cryptographic panorama. A technology once believed to lay far away into the future is increasingly closer to real world applications. Quantum computers will break the algorithms used in our public key infrastructure and in our key exchange protocols, forcing a complete retooling of the cryptography as we know it. Quantum Key distribution is a physical layer technology immune to quantum or classical computational threats. However, it requires a physical substrate, and optical fiber has been the usual choice. Most of the time used just as a point to point link for the exclusive transport of the delicate quantum signals. Its integration in a real-world shared network has not been attempted so far. Here we show how the new programmable software network architectures, together with specially designed quantum systems can be used to produce a network that integrates classical and quantum communications, including management, in a single, production-level infrastructure. The network can also incorporate new quantum-safe algorithms and use the existing security protocols, thus bridging the gap between today's network security and the quantum-safe network of the future. This can be done in an evolutionary way, without zero-day migrations and the corresponding upfront costs. We also present how the technologies have been deployed in practice using a production network.**

## I. Introduction

Security has always been a concern when speaking about critical information and infrastructures. Vulnerabilities can compromise entire networks at all levels, causing not only large economical losses and breaches of personal rights or loss of critical information, but also damage the public image and erode the confidence of users. The advent of quantum technologies, more particularly quantum computing, has brought new security threats which affect most of the existing security protocols. These protocols rely on mathematical problems that are computationally complex using traditional computers. However, quantum computing has demonstrated to be capable of breaking some of the most used. The mathematical basis of algorithms like RSA and ECC are weak under quantum attacks and so are other common protocols that use them, like Diffie-Hellman. The advancements of quantum technologies are bringing quantum computers closer in time. Devices once believed to be several decades in the future are now forecasted to happen within the next five years [1]. Albeit predictions are always risky, the trend is clear. Since the information can be recorded now to be deciphered later, any communication that is required to remain secret for a few years is potentially in danger today if it is protected only with these algorithms. This situation forced institutions such as the US National Security Agency to deprecate the current public key algorithms used in their communications, and start substituting them with quantum-safe methodologies as early as 2015. Other organizations, in particular the ones related to standardization, like the NIST, ETSI, IEEE, IETF, ISO/IEC and ITU-T followed suit [2]. Since changing the security infrastructure would require years, it is of fundamental importance to start thinking about a future-proof security infrastructure today.

There are two avenues to create this quantum-safe infrastructure. One is to substitute the current algorithms for others that could resist the attacks of a quantum computer and the other is to use quantum cryptography. The former relies on algorithmic complexity, while the latter is a physical layer technology immune to any computational threat. The standardization of both have already started [3,4] but it will take time, first to agree the standards themselves, and then to implement them at a global scale. In the case of post-quantum algorithms and according to NIST, the time frame would easily be well within the predicted danger zone. It is thus important to speed the adoption of quantum-safe technologies, especially in critical infrastructures.

Quantum cryptography [5] most prominent and mature result is Quantum Key Distribution [6]. The QKD technology can be described as a way to provide two synchronized sources of random bits, separated in space and with the property that the information about the bits leaked outside of the pair can be limited as much as desired. This allows to create a secret symmetric key between the pair. While post-quantum algorithms, still a matter subject to intense research, will be always dependent on the computational power ascribed to the adversary, QKD can be mathematically proven to be secure, independently of the resources of the adversary: It is an information theoretic secure -ITS- primitive [7]. Also, QKD keys are not correlated in any way among them, thus guaranteeing forward and backward security.

However, the physical layer restrictions and the point-to-point nature of QKD has lead this technology to be implemented in ad-hoc and restricted (usually point-to-point) scenarios. Even the most advanced demonstrators show QKD networks (trusted-relay-based) not integrated in network management systems, leading to non-optimal implementations. This actually amounts to impose the building of separate ad-hoc networks for QKD purposes. This is extremely expensive and cumbersome, explaining why QKD technologies, and therefore QKD networks, are far from being broadly adopted in network infrastructures.

The use of Telco networks is the only way to push forward QKD technologies for communication services. First applications will be deployed for corporate environments, which show more security related concerns. This should enable economies of scale that will eventually lead to generalize these solutions on residential services.

The integration of software defined networking (SDN) technologies, through the development of standard protocols and interfaces, has allowed new services and systems to be seamlessly integrated in telecommunications networks. It allows to manage and optimize the entire infrastructure from a central platform, the SDN controller. The flexibility brought by SDN reduces drastically the effort of integrating new devices and technologies in the network. In particular, the success of QKD networks radically depends on the degree in which they can be adopted in the existing infrastructure. The flexibility of the Software Defined Networks (SDN) allows the integration of quantum communications in the telecommunications network to a degree that was simply impossible in previous schemes, where network devices should be modified, one by one, to create a quantum channel.

The objective of this work is to demonstrate, for the first time, a network that can seamlessly integrate, in a logical and physical way, quantum and classical communications in a production level telecommunications infrastructure. The scheme is also compatible with existing security protocols as well as post-quantum ones, allowing for an evolutionary path to quantum-safe infrastructures.

This work is organized as follows: Section II presents the QKD characteristics and network approaches used until now; Section III proposes a high-level abstraction model for software-defined QKD nodes; Section IV describes the testbed network used for the implementation, while Section V presents the conclusions.

## II. QKD Technology and Networks

QKD is a physical technology. It allows the growing of a secret key that is shared only by the legitimate users of a quantum channel connecting them. This channel is usually implemented using optical fiber as a transport media for the quantum signals (qubits). Thus, QKD can be seen as a technology to extend the security perimeter to the optical fiber used to link them. To perform QKD, a public and authenticated classical channel is needed in addition to the quantum channel, in particular for post-processing operations, such as key distillation and privacy amplification. During the installation process of a new QKD device, an initial secret is assumed to be shared among both ends -customarily named Alice for the emitter and Bob for the receiver- for the first authentication. Further authentication rounds between Alice and Bob (not just in a direct link, but also in a multi-hop QKD link) are done using the freshly generated key material. Assuming a correct implementation, installation and no side channels, QKD can be demonstrated to be absolutely secure. This means that the information on the secret key leaked outside the emitter/receiver pair can be made arbitrarily small. Unlike conventional algorithms, QKD is not based on computational assumptions, like the complexity of certain mathematical problems, but on the laws of Physics. This makes it resistant to any computational advance, like quantum computing.

However, this physical nature makes QKD intrinsically distance-limited, as qubits have a finite probability to interact with the transport medium. A signal propagating in a medium suffers an exponential attenuation and this is critical when they are a single quantum. Also, the interactions with the environment are indistinguishable from the action of a spy and errors must be treated as if they were the action of an eavesdropper. This heavily penalizes the secret key throughput. As a typical example, in the case of qubits implemented in photons and using optical fiber, losses are about 0.2 dB per Km when using the telecommunications C-band at 1550 nm. QKD systems working with losses of about 30 dB producing a significant key rate have been demonstrated, and technology is reaching maturity quickly. This means that today’s practical limit in distance is about 150 Km just considering the fiber attenuation. Going beyond means a strongly reduced secret key rate output. Top performance figures are about 1Mbps of final secret key at 40 Km distance in direct links, without crossing any passive network equipment that would increase losses and penalize the throughput. This distance limits could be eventually overcome by using quantum repeaters, which would also allow other uses, but this is a technology still years in the future [8].

As a consequence, the current security model of a QKD network is that of a collection of interconnected trusted nodes, a structure that fits well with modern optical infrastructures, where QKD can be used to connect the security perimeters protecting the network points of presence where the control plane instances are running. For network operators, inter datacenter as well as last mile deployments to business customers are the scenarios where QKD would apply. In this sense, QKD is a realistic technology to build a physical security layer to protect the infrastructure [9] and network services [10,11]. The importance of the testbed used in this communication lies in that it demonstrates how QKD can be implemented in a real-world network, installed in a production facility and run in a fully integrated manner, where the quantum and classical parts are managed consistently. This allows the incremental installation of QKD and QKD-related services, avoiding the use of a separate network for the quantum channels and the large up-front costs associated, as has been done in the past. Moreover, the testbed has been used to show industrially relevant use cases [9,11] for service and infrastructure security.

QKD links generate key material at the two endpoints of the quantum channel (Alice and Bob), but different strategies have been designed to enable key delivery among multiple points connected via quantum links. For longer distances than the ones supported by a single link, the key material is distributed using secured and trusted intermediate nodes, all connected by QKD links composing a network. In order to securely forward the keys, each hop/link must provide ITS encryption and authentication.

This approach has been demonstrated to be compatible with metropolitan area networks, either running exclusive quantum networks, in parallel with operator's infrastructures, notably the pioneering DARPA network [12], or to explore the capability to run in coexistence with classical communications [13]. More specifically, the last ones focus on direct switched connections using as much off-the-shelf optical components as possible. Moreover, when speaking about QKD network management, the most mature developments come from the SECOQC and Tokyo projects [14,15], both essentially with the same structure. In SECOQC they developed a logical layer to enable key delivery and secure channels among trusted locations within the QKD network. This key distribution management layer was implemented following traditional networking principles. A point-to-point protocol provided a mechanism to make any QKD network location aware of the state of the quantum network (e.g. QKD network nodes and their connectivity, key availability per link), similar to other existing routing protocols, like the open shortest path first -OSPF-. In this way, each node was responsible for accepting or denying new requests, as well as to decide which path is the appropriate to instantiate new multi-hop connections. This might lead to non-optimal strategies and a misuse of the key material. Such optimization and synchronization issues can be mitigated when centralizing the network planning. The SDN control plane could not only dynamically plan the creation of multi-hop QKD key delivery across a network, but also automate the physical link creation and the installation process when adding new QKD systems. Crucially, it can also manage the coexistence of quantum and classical channels sharing the same network. This work introduces new directions for the integration of QKD resources into novel networking architectures, in the form of new abstraction models, architectures and use cases.

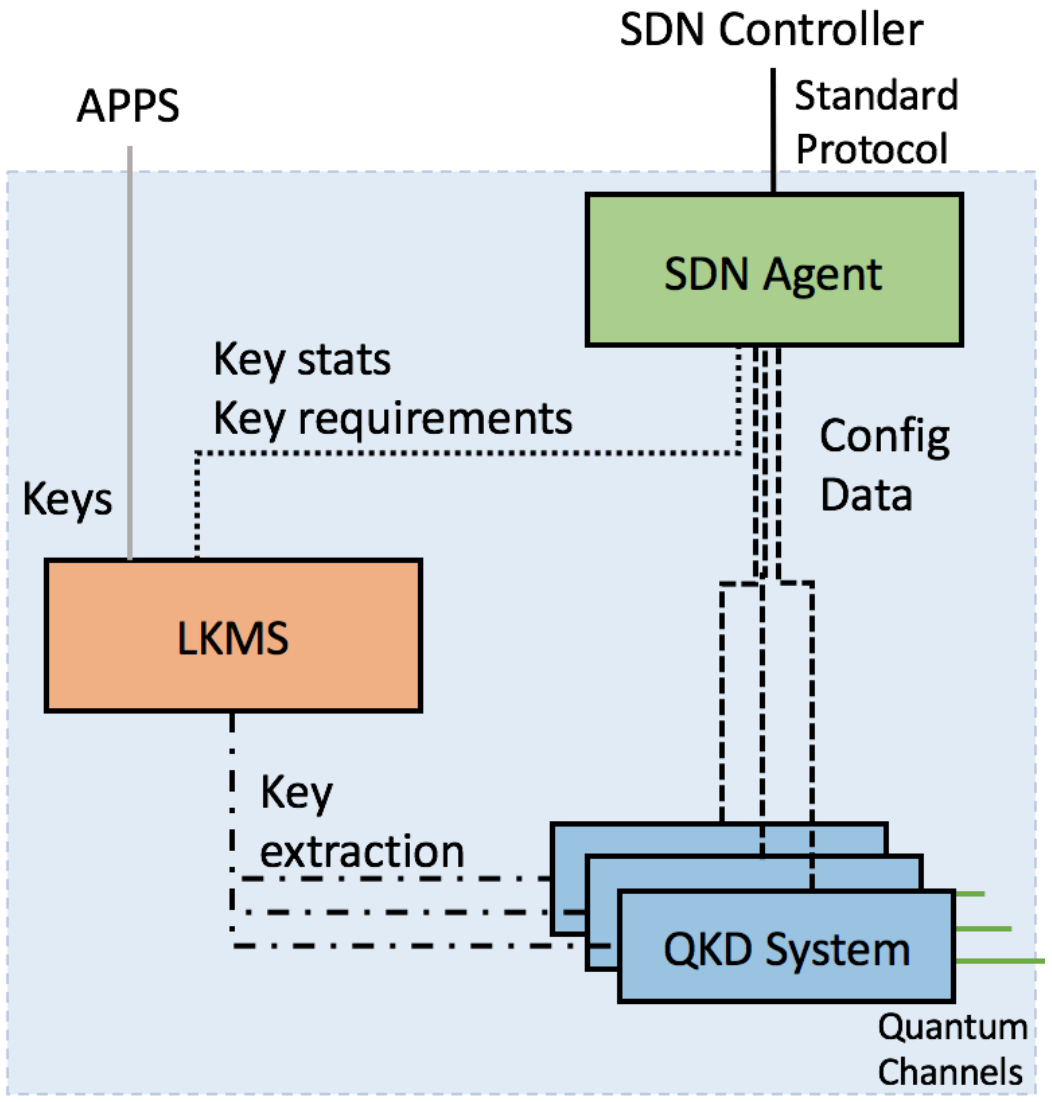

**Figure 1:** Definition of a Software Defined QKD Node. One or many QKD systems work under the control of a SDN agent that is in charge of communicating with a central SDN controller. The products of the QKD systems (keys) are managed with the help of an entity referred as the Local Key Management System (LKMS, see text.).

## III. Abstraction Model for Software-Defined QKD Nodes

The main idea behind the proposed scheme is to model any type of QKD system as similarly as possible to any other network elements (e.g. router, switch) so a SDN controller will be capable of understanding, integrating and optimizing their behavior. Creating such model requires an analysis of which capabilities and parameters from the QKD domain are useful from the network management perspective. For instance, QKD systems are highly complex, have multiple operational hardware and software subsystems whose complexity should be hidden from the network controller. Also, other data produced and maintained by the QKD systems, such as internal information or the actual key material, must be kept confidential, as they are not required for network operation and their exposure could compromise the security of the QKD network.

This model serves different purposes: firstly, moving towards standard interfaces will allow to reduce the time-to-market of a solution whose major hard-stopper is being stuck to point-to-point ad-hoc implementations; secondly, the advent of SDN technologies for optical networks facilitates the spectrum allocation for the quantum channel; and finally, the centralization of the QKD network management will allow to better optimize generation and utilization of QKD-derived keys, while maintaining a central database of active links and consumer applications.

Our proposal is then to aggregate and abstract any QKD system within the same secure area as interfaces of a logical network element. Let us describe the node architecture depicted in Figure 1: we consider the whole set of elements as part of our Software-Defined QKD (SDQKD) node. Down at the right bottom of the Figure 1, we can see several QKD systems under the same location and logical control. These physical systems are in charge of creating quantum channels and generating

QKD-derived keys, which are then stored at the local key management system (LKMS). The LKMS maintains the keys generated from the different QKD systems and gathered via a key extraction/delivery interface using an interface specification from the ETSI ISG-QKD group.

Finally, the SDN agent is capable of gathering important data from the node, communicating with the SDN controller and handling and activating the configuration updates requested by the controller. These operations will require the agent to orchestrate the other elements within the node: the LKMS and the QKD systems.

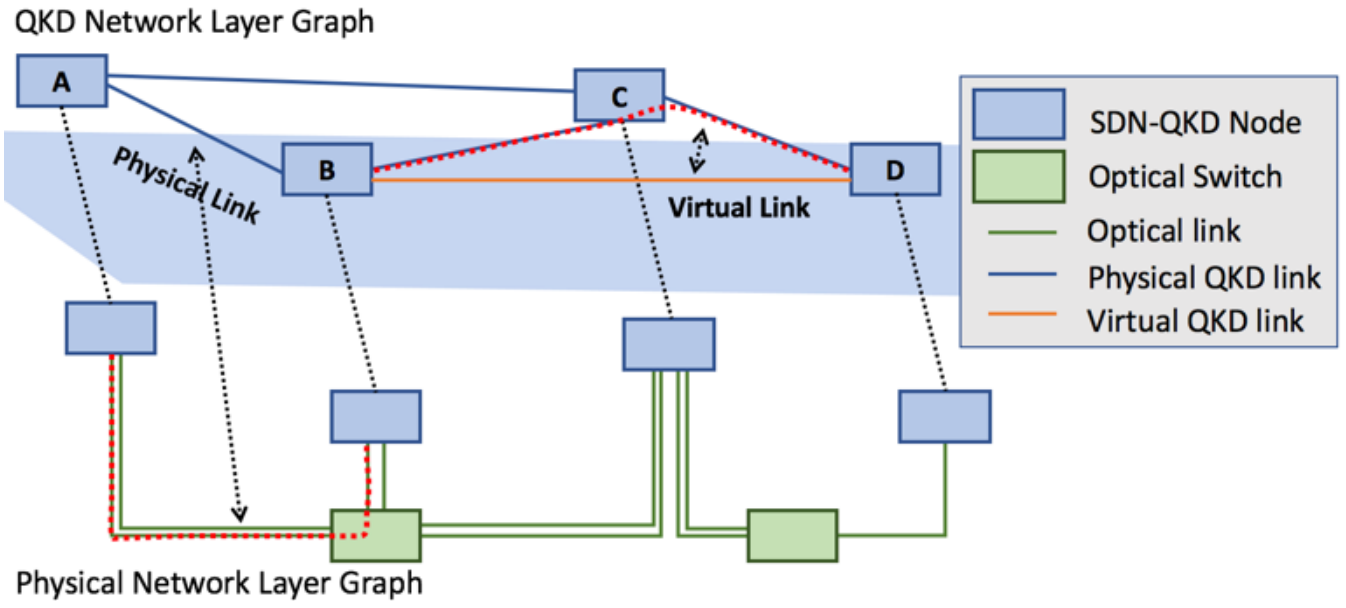

**Figure 2:** Physical and logical view of a SDN QKD network. A layer that contains the physically available network is used to create the functional QKD network. Note the virtual link that connects SDQKD nodes B and D through a multi-hop connection that passes through SDQKD node C.

In our design, the SDQKD node model is structured in four key components:

- *Interfaces*: all the QKD systems aggregated under a SDQKD node.
- *Applications*: any entity, either external or internal, using keys from the node's LKMS.
- *Links*: considered as "key" associations between two SDQKDNs, either physically connected via a quantum channel, or based on trusted-relay –virtual– schemes.
- *Capabilities and additional information*: additional data, such as identifiers and locations, and support for multiple operations.

This abstraction model is on a definition phase within the ETSI industry specification group (ISG) for QKD. While exploring every single parameter of the model is out of the scope of this work, describing the use cases supported by the SDQKD paradigm and model while looking at Figure 2 will provide a better overview of the solution. Some of the fundamental use cases which are the base for our SDQKD network are:

- *Creation of physical QKD link*: a physical QKD link is, in other words, a pair of QKD systems connected via a quantum channel capable of generating keys (in Figure 2, the link between A and B). A quantum channel is carried by a point-to-point optical connection with no intermediate active components or regeneration. This process can be automated by the controller to optimize the spectrum allocation and to monitor the performance of the link.
- *Creation of a virtual QKD link*: a virtual link is an association between two SDQKD nodes which generates keys through intermediate trusted nodes forwarding key material using ITS procedures, also known as trusted-relay mode. To keep the same security level, OTP is used for the key relay. In Figure 2, the link between B and D, via trusted-node C.
- *Application management*: The SDN controller, via notifications and other monitoring messages, is responsible for detecting and managing client applications. In this way, applications do not need to be aware of the QKD network, but just of their respective endpoints for requesting key material. This also allows to keep a central repository of applications and to monitor their key usage for accounting purposes.

These three base operations form the core for more complex management and use cases (e.g. to create multicast associations for securing sessions shared between multiple parties) and are the primary focus of the Madrid's SDQKD network demonstrator, presented in the next section.

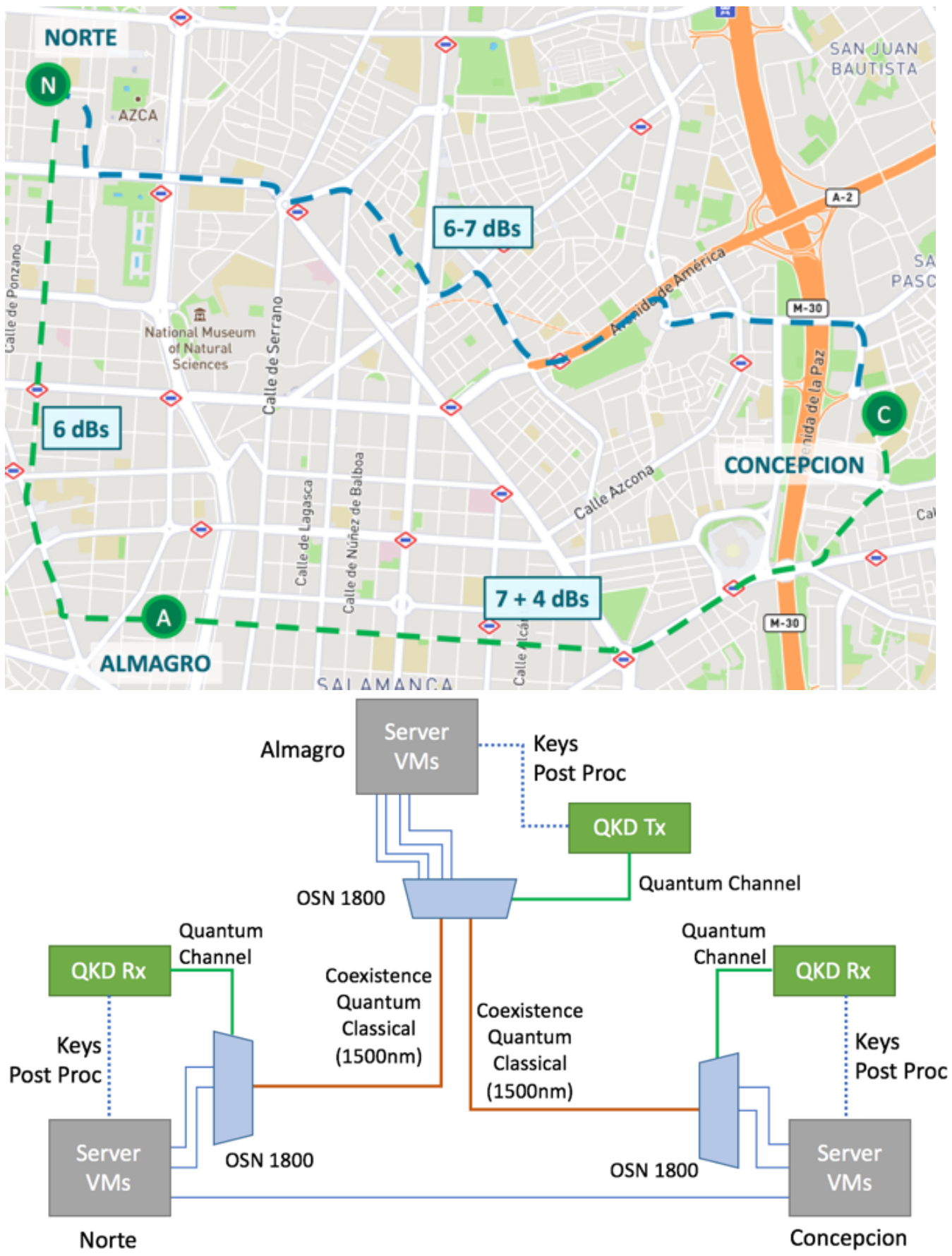

**Figure 3:** Map of the Madrid SDQKD network and logical representation of the elements involved in the field trial. Three Telefonica's production facilities are connected in a ring.

## IV. TESTBED

To show the technological maturity of these concepts, we have implemented a first-of-a-kind SDQKD network in Madrid. The uniqueness of this network is based on several reasons:

- Field-trial on a production network: It is the first demonstrator using a commercial production network. The

fibers connect production points of presence (PoPs) of Telefonica Spain's metro network in Madrid. No difference was made in the procedures of the installation of the standard telecommunications equipment and the QKD devices, highlighting their stability and robustness.

- Continuous variable QKD (CV-QKD): The QKD technology used in the trial is CV-QKD, which is known to be more resilient to noise. The equipment, developed by the Huawei Quantum Communications Laboratory in Huawei Research Germany, has been designed considering the SDN principle, as software driven and programmable systems. Although the network is implemented using this technology, the SDQKDN model definition is designed to cope with any existing QKD technology and vendor system.
- Quantum channel co-propagation: The quantum channel is co-propagated with commercial classical communications. The quantum channel is capable of tolerating more than 20 classical channels at 100Gbps in the same C band than the quantum channel. The testbed uses commercial optical transport network (OTN) platforms. The quantum channel is injected in the network using these same devices, without requiring any non-standard equipment.
- SDN managing the QKD resources: It is the first SDN managed QKD network in a production environment. The creation of links (physical or virtual) and the management of client applications is done via standard protocols and a well-defined model that includes the necessary parameters to manage and optimize the quantum network.
- QKD securing the control plane: QKD is added in an incremental way to the existing security infrastructure: whenever composable, keys obtained using QKD and standard protocols are XOR-ed, so that the certifications of the existing security protocols remain applicable. This means that the QKD layer can be seen as a way to seamlessly upgrade the security to a quantum-safe grade, without destroying the trust in the current classical mechanisms. Note that this is also an easy way to incorporate post-quantum algorithms.
- Additional telecommunications use cases: Furthermore, different use cases ([9,11], among others to be published) were implemented on top of our SDQKD network to secure multiple communication layers among PoPs.

The testbed connects three Telefonica Spain PoPs with three fiber pairs that are part of Telefonica Spain's production network: Almagro (hosting a Telefonica I+D/gCTIO laboratory shared with a Telefonica Spain PoP), Norte and Concepcion. The fibers go across other elements that are part of the commercial network, including patch panels and other connectors. They cause losses on the optical channel higher than what can be expected from the distance alone. Almagro hosts the QKD transmitter, while Norte and Concepcion host the two receivers. The QKD network, following SDN principles, is optimized in such a way that the single transmitter can generate keys with the two receivers having minimum (even none) performance penalties, since while the transmitter generates keys with one end, the other uses this interval to perform its required calibration process. Some general parameters of the links:

- Almagro-Norte: 3.9 Km length and around 6dBs of losses, with a key generation rate calculated from the measured signals above 70kbps, with coexistence of quantum and classical channels.
- Almagro-Concepcion: 6.4 Km length and around 7 + 4dBs of losses (the 4dBs, coming from an additional 20Km fiber spool to test longer distances), with a key generation rate above 20kbps, with coexistence of quantum and classical channels.
- Norte-Concepcion: 5.5 Km length and 7 dBs of losses, carrying just classical channels. This link is used to demonstrate a multi-hop/virtual link using SDN and QKD keys.

Figure 3 depicts the logical and physical maps of the Madrid SDQKD network. In the logical schema (Fig. 3, lower part) the devices at each location are shown. We can find, apart from other devices for measuring and gathering results, three fundamental systems: the classical optical communication systems (the abovementioned OTN platforms), the CV-QKD systems and the servers which include the management software, the signal post-processing and the different use cases/experiments carried out during the field trial. Multiple interfaces from the server are connected to the OSN to separate the different channels (control, for QKD and network, and data) and multiplexed together with the quantum channel. Norte and Concepción servers where directly connected via dedicated fiber, as there was no direct quantum channel between those two locations. Regarding the logical systems, both controller and agents implements RESTCONF interfaces to communicate between each other. They are based on the abstraction logic presented above written in YANG, which is extensible and can adapt to the specific capabilities of the QKD device. Additionally, the controller hosts a RabbitMQ server for notifying changes from the SDQDKN agents, such as new applications connecting to the nodes."

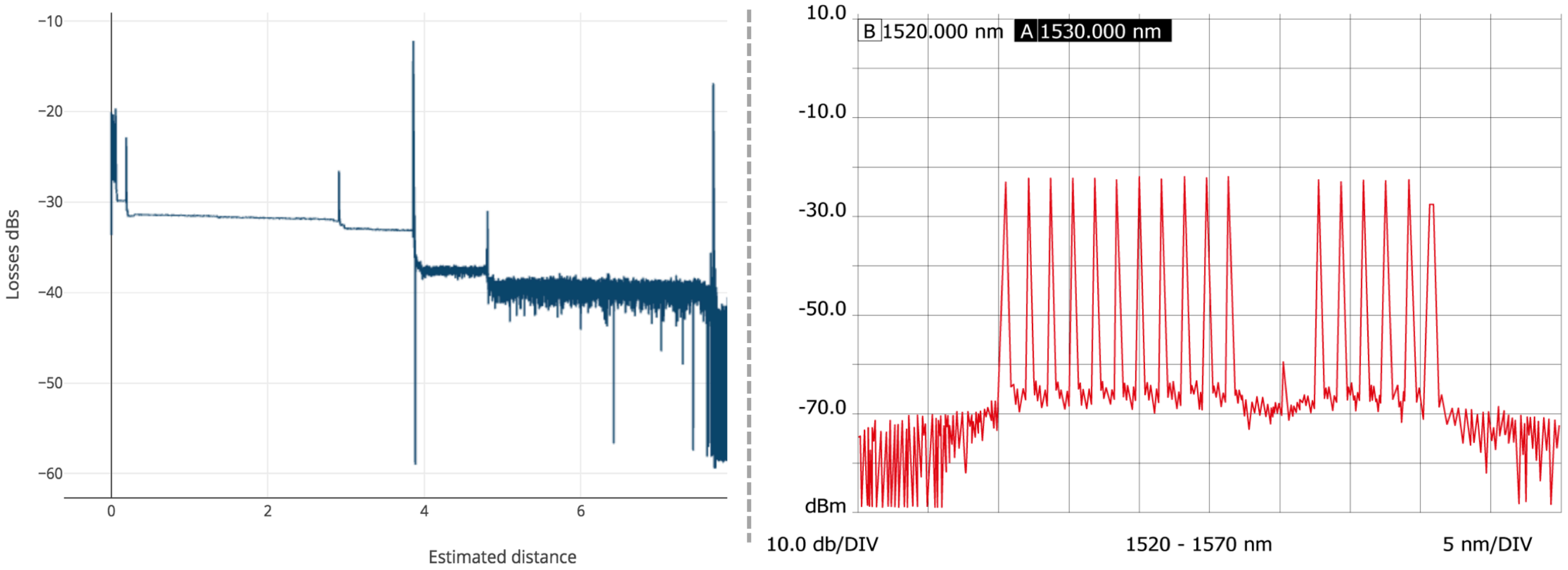

**Figure 4:** (left) OTDR measurement between the Almagro and Norte facilities showing the distances and losses. (right) OSA capture of the link between Almagro and Norte carrying the quantum and classical communication signals. The large spikes correspond to the classical channels (measured at the receiving end. Their power was more than enough to have classical communications without increasing the typical BER). The small spike at about 1550 nm corresponds to a pilot tone used to mark the position of the quantum channel.

To show an example of the characteristics of the optical channels, Figure 4 displays the OTDR and OSA measurements of one of the links (from Almagro to Norte). The OTDR measurement shows both directions of the links, as a loopback connection was stablished during the installation.

The losses in this link are around 6 dBs in each direction. The big spike at the middle of the graph at about 3.9 km represents the loopback connection at Norte, while the other two spikes shows other passive components at an intermediate PoP that does not belong to the QKD network.

The OSA capture shows the classical channels coexisting with the quantum channel. In the testbed, up to 17 classical channels were used (See Fig. 4, right panel), the maximum being dictated by the number of available physical ports The pilot tone located between the classical channels 11 and 12 (approx. 1550 nm) is used to locate the quantum channel and for synchronization purposes.

## V. Conclusions and Future Challenges

As any technology still in the applied research phase, QKD has to go through several stages to reach the necessary maturity, also for its certification and standardization, before reaching market-ready status. Apart from the well-known issues related to hardware interoperability issues (common also in traditional networks), architectures for any QKD network have to be agreed and designed. These designs should fit any implementation, either for QKD networks coexisting with conventional communications or for separated networks. They also should foresee and be suitable for scenarios with dynamic quantum channel allocation based on implementation of SDN, and fixed (not reconfigurable) schemes.

The demonstrations reported in this paper presents a clear step forward in enabling the integration of QKD systems in current network operator infrastructures. We have used SDN principles to produce a converged quantum-classical network that shares the physical and logical infrastructure among quantum and classical channels. In contrast to the typical quantum networks architectures, that use an ad hoc network, running in parallel to the conventional for qubit transmission, the quantum-enabled SDN architecture proposed and demonstrated here unites under the same management the quantum and classical communications, enabling network optimization to better use all resources. The developed network supports the existing security mechanisms and avoids large upfront deployment costs by allowing for staged deployment. The demonstration in production sites showcases the maturity of the technology for real world services. This paves the way for a broader uptake of quantum technologies as a solid building block for the quantum-safe networks that will be needed in the future.

## Acknowledgements

We thank the support of projects CVQuCo: MINECO/FEDER TEC2015-70406-R, CiViQ: European Union's Horizon 2020 research and innovation programme under grant agreement No. 820466 and QUITEMAD: Comunidad de Madrid P2018/TCS-4342.

**Alejandro Aguado** received the Graduate degree in mathematics and computer science from the Universidad Autonoma de Madrid, Madrid, Spain, in 2014. He worked as a researcher on SDN with Telefonica I+D. He worked also as a Research Associate at the High Performance Networks Group, University of Bristol. He is currently a PhD student in the Center for Computational Simulation, Universidad Politécnica de Madrid, researching on quantum key distribution networking.

**Victor López** M.Sc. from Universidad de Alcalá de Henares (2005) and Ph.D. from UAM (2009). In 2006 he joined the High-Performance Computing and Networking Research Group, UAM, where he worked as an assistant professor. In 2011 he joined Telefonica I+D as a technology specialist. He has co-authored more than 150 publications, 4 patents and contributed to IETF and ONF. His research interests include IP/optical networks and their programmability (SDN, PCEP). He's editor of the book Elastic Optical Networks: "Architectures, Technologies, and Control".

**Diego López** M.S. from the University of Granada (1985) and Ph.D. dfrom the University of Seville (2001). He joined Telefonica I+D in 2011 as a senior technology expert where leads the Technology Exploration activities within the GCTO Unit. He is focused on network virtualization, infrastructural services, network management, new network architectures, and network security. He participates in the ETSI ISG on NFV, chairing its Technical Steering Committee, the ONF, and the IETF WGs on these activities

**Momtchil Peev** studied mathematics and physics at the University of Sofia, Bulgaria. Ph.D. in solid state physics in 1993. From 1993 to 1995, he was a post-doctoral Lise-Meitner Fellow at the Vienna University of Technology and, from 1995 to 1997, a post-doctoral ARCS Fellow. Research associate at ARCS (resp. AIT) until 2010, when he became senior scientist and a thematic coordinator for QKD. Since 2015 he is a senior expert and project leader in the Optical and Quantum Communications Laboratory at the Munich Research Center of Huawei Technologies Duesseldorf GmbH. M. Peev led the development of a QKD network prototype in the European project SECOQC (2004-2008), and coordinated the effort of the Huawei Munich Group in the SDN-QKD demonstration of Telefonica-UPM-Huawei in Madrid (2018).

**Andreas Poppe** studied Telecommunication Engineering at the Vienna University of Technology, Austria. He received his Ph.D. degree at the Photonics Institute in 2000. He joined the quantum cryptography project at the Institute of Experimental Physics (U. Vienna), where he designed an entangled QKD-system, included in the QKD-network demonstration of the European project SECOQC. He was senior scientist at the Optical Quantum Technologies group of the Austrian Institute of Technology, where he led the QKD systems development team. In 2015 he joined the Optical and Quantum Communications Laboratory at the Munich Research Center of Huawei Technologies Duesseldorf GmbH as a senior expert.

**Antonio Pastor Perales** is an Industrial Engineer from the Carlos III University of Madrid (1999). He joined Telefónica I+D in 1999 working on the design and deployment of different networks worldwide. Since 2006 he has been working as an expert in network security in Telefónica. He is currently the expert responsible for security research in the network transport area in Telefonica Global CTIO Unit. He has participated in many security research programs, holding several security certifications and patents.

**Jesús Folgueira** MSc in Telecommunications Engineering from UPM (1994) and MSc in Telecommunication Economics from UNED (2015). He joined Telefónica I+D in 1997. Head of Transport and IP Networks within Telefonica Global CTO unit, he leads Network Planning, Technology and Innovation. He is focused on Optical, Metro & IP Networks, network virtualization (SDN/NFV) and advanced switching. His expertise includes Broadband Access, R&D Management, and network deployment.

**Vicente Martín** Ph.D. Physics (1995) from the Universidad Autónoma de Madrid. Full Professor at Universidad Politécnica de Madrid (UPM). Founding member of the Specialized Group in Quantum Information and Computing of the Spanish Royal Society of Physics and of the Quantum Industry Specification Group at the European Telecommunications Standards Institute. His main research interest is in quantum cryptography and its integration in conventional networks. He leads the Research group on Quantum Information and Computation Group at UPM and is also director of Center for Computational Simulation.